\documentstyle{article}    %for ordinary Latex, delete if laa.sty available
\newcommand{\inst}[1]{ (#1)}

\newcommand{\sun}{\mbox{$\odot$}}
\newcommand{\thesaurus}[1]{}
\newcommand{\offprints}[1]{\footnotetext[1]{{\it Send offprints
requests to} #1}}
\renewcommand{\and}{\\}
\newcommand{\institute}[1]{\footnotetext[2]{Address:\\ #1}}
\newcommand{\la}{\le}
\newcommand{\ga}{\ge}
\newcommand{\keywords}{{\bf Keywords:} }
\newcommand{\arcsec}{\mbox{''}}
\newcommand{\picplace}[1]{\frame{\centerline{Figure}}}
\newcommand{\acknowledgements}{{\it Acknowledgements.}}
\renewenvironment{thebibliography}[1]{{\section*{References}}
\parindent0cm(in alphabetical order)}{\\}
\renewcommand{\bibitem}[7]{\\}
\newcommand{\AGNi}{Active Galactic Nuclei{ } }
\newcommand{\GC}{Galactic Center{ } }

\newcommand{\smlcps}{\sc}
\begin{document}
%%%%%%%%%%%%%%%%%%%%%%%%%%%%%%%%%%%%%%%%%%%%%%%%%%%%%%%%%%%%%%%%%%%%%

\thesaurus{04(02.01.2; 02.02.1; 10.103.1)}
\title{A rotating black~hole in the Galactic Center}
\author{
Heino Falcke \inst{1}
\and Peter L. Biermann \inst{1}
\and Wolfgang J. Duschl \inst{2}
\and Peter G. Mezger \inst{1}}
\offprints{HFALCKE@mpifr-bonn.mpg.de}
\institute{Max-Planck-Institut f\"ur Radioastronomie, Auf den H\"ugel 69,
D--W5300 Bonn
\and Institut f\"ur Theoretische Astrophysik der Universit\"at Heidelberg,
Im Neuenheimer Feld 561, D--W6900 Heidelberg}
\date{Accepted for publication in A\&A: December 4, 1992}
\maketitle
\begin{abstract}
Recent observations of Sgr~A* give strong constraints for possible
models of the physical nature of Sgr~A* and suggest the presence of a
massive black~hole with $M\la 2\cdot 10^6 M_{\sun}$ surrounded by
an accretion disk which we estimate to radiate at a luminosity of $<
7\cdot 10^5 L_{\sun}$. We therefore calculate the appearance of a
standard accretion disk around a Kerr hole in Sgr~A* following from
general relativity and a few fundamental assumptions. Effective
temperature and luminosity of the disk spectra do not depend on the
unknown viscosity mechanism but instead are quite sensitive to
variations of intrinsic parameters: the mass, the accretion rate, the
angular momentum of the accreting hole and the inclination angle.

A radiation field of $L\simeq 7\cdot 10^4 - 7\cdot 10^5 L_{\sun}$ and
\mbox{$T_{\rm eff}\simeq$} \mbox{$2-4\cdot 10^4 {\rm K}$} can be ascribed to
a rapidly
rotating Kerr~hole ($a>0.9$) accreting $10^{-8.5} M_{\sun}/{\rm yr}
<\dot{M}<10^{-7} M_{\sun}/{\rm yr}$ at a black~hole mass of $M\la
2\cdot 10^6\,{\rm M_{\sun}}$ seen almost edge on.  A low mass
black hole of $M\la 10^3 M_{\sun}$ seems to be very unlikely.

 Due to the large uncertainties in the observational determination of
the effective temperature further obervations are required.  Therefore
we provide a ``Hertzsprung-Russell diagram for black~holes'' together
with simple scaling laws to provide a test for the black
hole/accretion disk scenario in the Galactic Center and give a direct
method to measure such intrinsic parameters as the angular momentum
and the accretion rate of the hole.

\keywords{Accretion disks -- black~hole physics -- Galaxy: center}
\end{abstract}

\section{Introduction}
There is mounting evidence for the presence of a black~hole surrounded
by an accretion disk at the center of our Galaxy, provided by the
following observations:

\begin{itemize}
\item[i)] The kinematics of gas and stars in the inner parsecs requires an
enclosed mass of a few $10^6 M_{\sun}$ within a galacto-centric
radius of $R \la 0.1\, {\rm pc}$ (see Genzel \& Townes 1987, Sellgren 1988
and references therein).

\item[ii)] Morphology and kinematics of the layer of interstellar matter
between the inner Lindblad resonance (i.e. $R\la 2.5\, {\rm pc}$) and
$R\simeq 1-0.1\, {\rm pc}$ show the expected characteristics of
accretion disks in \AGNi (AGN) (Zylka, Mezger \& Lesch 1992 (hereafter
ZML92), and references therein)

\item[iii)] Close to, or at the dynamical center of the Galaxy is the
variable radio source Sgr~A* interpreted as synchroton emission. Its
mm/submm and FIR spectrum is best explained by $\simeq 30-60 {\rm K}$
dust emission from an irradiated accretion disk (ZML92) which
extends to $10^{17} {\rm cm}$ and contains a mass of $400 - 800 M_{\sun}$.

\item[iv)] The luminosity of the compact central object is $L\la
7\cdot 10^5 L_{\sun}$, far below the Eddington limit of $\simeq
2-4\cdot 10^{10}L_{\sun}$ for a black~hole with $M\simeq
1-2\cdot 10^6 M_{\sun}$. A cavity in the Sgr~A East core could be due
to an explosive event in the past, when this accretion rate was
several magnitudes higher (Zylka et al 1990).

 \item[v)] The radio emission of Sgr~A* could be explained as emission
from a small scale nuclear jet, stressing the analogy to AGN (Falcke,
Mannheim \& Biermann 1992) where a massive black~hole with an
accretion disk is still the best explanation for the central engine.
7~mm (Krichbaum et al. 1992) and 3~mm VLBI observations of Sgr~A* --
currently being evaluated or prepared -- will therefore give a firm
test for the whole scenario.
\end{itemize}

Hence if there is a black~hole surrounded by a hot accretion disk in
the center of the Galaxy it would be underfed at present with an
accretion rate of $\dot{M}\ll 10^{-6} M_{\sun}/{\rm yr}$ as compared
with an accretion rate of $\simeq 0.03 M_{\sun}/{\rm yr}$ in the
Eddington limit. The luminosity of $\la 7\cdot 10^5 L_{\sun}$ of the
starved black~hole would be mainly contributed by thermal free-free emission
from a $T_{\rm e}\simeq 2-4\cdot 10^4 {\rm K}$ plasma at the inner
edge of an accretion disk (ZML92) in agreement with recent NIR
observations by Eckart et al. (1992).

Various other attempts fail to explain these features of Sgr~A*.
Therefore in this paper we propose to test the black~hole concept in
more detail by calculating the appearance of a disk around a rotating
black~hole, considering specifically the effects of the Kerr metric on
the spectrum.  We will demonstrate that a black~hole mass of about $2
\cdot 10^6 \,{\rm M_{\sun}}$ and the radiation field described above
require a strongly rotating Kerr black~hole with a low accretion rate
and seen almost edge on.

However, at the moment the obervational basis is too weak to give an
unambiguous answer to the question whether such a configuration exists
in the Galactic Center, since most of these observations are at the
limit of what is technically possible. For an observer it should
therefore be of highest interest to know, which the crucial
observations are for a confirmation of a black~hole/accretion disk
configuration and for a determination of its parameters.  We therefore
compute in this paper the characteristics of such a configuration for
a parameter range suggested by the observations above. We find that a
refined measurement of the effective temperature and the luminosity of
the central object together with our grid of parameters should make it
possible to differentiate between parameter combinations for a black
hole/accretion disk model of the \GC and even gives an estimate for
the angular momentum of the hole.

\section{A critical review of the crucial observations}
The crucial observations related to the nature of the central compact
object and its surrounding disk are: the kinematics of gas and stars,
the mm-to-MIR spectrum of the central arcseconds, the NIR spectrum of
Sgr~A* and its effective temperature and luminosity.

Genzel \& Townes (1987) and Sellgren (1988) -- based on the same
observational results -- arrive at somewhat contradictory
interpretations. Genzel \& Townes state that the data are best
approximated by a star cluster and a central point mass of $\simeq 2.5
-3\cdot 10^6 M_{\sun}$. Sellgren states that -- if the core-radius of
the star cluster is $\la 0.1\, {\rm pc}$ -- no central point mass is
required, although in our opinion the model fit shown in her Fig. 4 requires
 a point mass of the order of $\sim 10^6 M_{\sun}$.

The interpretation of the mm-to-MIR spectrum in terms of dust emission
(ZML92) from a compact cloud of $\simeq 1\;{\rm arcsec}$ diameter
could in principle be strengthened by interferometric observations at
$\lambda 1 {\rm mm}$ and by additional flux density measurements in
the $\lambda 460\mu{\rm m}$ and $\lambda 350 \mu {\rm m}$ windows.
These observations have been made but are not yet evaluated.

How well are the effective temperature and luminosity of the central
object known? The dereddened spectrum (Eckart et al. (1991) adopt
$A_K=3.4$ and  $A_H=5.4$ corresponding to $A_v=27$ mag) can be
approximated by a Rayleigh-Jeans spectrum. Hence, if the radiation is
thermal free-free emission from an opaque plasma, its electron temperature
must be $T_{\rm e}> 10^4 {\rm K}$.  This sets a lower limit for both
 $T_{\rm e}$ and $L$. Combining the Rayleigh-Jeans approximation $S_{\rm
NIR}\propto \Omega_{\rm s} T_{\rm e}$ ($\Omega_{\rm s}$ is the source
solid angle) with the integrated Planck spectrum $L\propto \Omega_{\rm
s}T_{\rm e}^4$ yields\footnote{This is the relation given by Eckart et al.
 (1991)
for a disk distribution of the surface brightness which is more likely
than the gaussian distribution adopted by ZML92, yielding $L=2\cdot
10^5 T_{{\rm e},35} L_{\sun}$.}
\begin{equation}
\label{L2Teff}
L=7.5\cdot 10^4 (T_{\rm e}/20\,000 {\rm K})^3 L_{\sun}.
\end{equation}
ZML92 obtained the luminosity of the central $30\arcsec$ by integrating the
 observed spectrum
\begin{eqnarray} \label{LIR}
L_{\rm IR}(30\arcsec;\lambda\ga 10\mu {\rm m})&=&L_{\rm IR}({\rm star
 \;cluster})+L_{\rm IR}({\rm central\;object})\nonumber \\
 &\simeq& 1.5\cdot 10^6 L_{\sun}
\end{eqnarray}

This assumes that heating of the dust is due to
 contributions from both the star cluster and a central object. The
contribution from the star cluster can
 be estimated from IRAS observations (Cox \&
Laureijs 1989). Fig. 14 of Cox \& Mezger (1989) shows the accumulated
bolometric luminosities of the star cluster (derived by Sanders and
Lowinger (1972) from the $\lambda 2.2\mu {\rm m}$ surface brightness)
and IRAS. Within $10\le R_{\rm pc} \le 200$ both have the same
functional dependence $L\propto R^{1.2}_{\rm pc}$. Specifically, the
accumulated IRAS dust luminosity is
\begin{equation}
L_{\rm IRAS}(R)=1.5 \cdot 10^6 R_{\rm pc}^{1.2}L_{\sun}
\end{equation} yielding $L_{\rm IR}(30\arcsec;{\rm star\;
 cluster},\lambda\ga 10\mu{\rm m})\simeq 8\cdot 10^5 L_{\sun}$ and leading
 to $L_{\rm IR}({\rm central\; object})\simeq 7\cdot 10^5 L_{\sun}$.

On the assumption that all radiation from the inner disk is absorbed by
 dust in the central $30\arcsec$ we can substitute the above luminosity in
 Eq. (\ref{L2Teff}) and obtain an electron temperature $T_{\rm e}\simeq
 42\,000 {\rm K}$.  This is an upper limit since all
other effects -- e.g. a substantial contribution of the ``He {\smlcps I}
emission-line stars'' detected by Krabbe et al. (1991) in the inner
parsec or a probable increase of dust extinction beyond $A_v=27$ mag
-- tend to decrease $T_{\rm e}$ but increase $L$ for a given $T_{\rm
e}$.
Thus

\begin{equation} 20\,000 {\rm K} < T_{\rm e}< 42\,000 {\rm
K}\end{equation}
and
\begin{equation}7\cdot 10^4 L_{\sun} <L({\rm
central\; object})<7\cdot 10^5 L_{\sun}\end{equation}
seem to be
reasonable limits for $T_{\rm e}$ and $L$ of the central object.

Although our estimate of the effective temperature falls exactly in
the regime derived before from the ionization state of gas in the
inner 3 pc (Genzel \& Townes 1987, Sect.~3, Roberts et al. 1991) it is
doubtful whether one can use these estimates as observational
parameters to fit accretion disk models to the central object.
Firstly, one should expect that with such low luminosities the
ionization of the gas is dominated mainly by hot stars in the central
star cluster (Krabbe et al. 1991) and secondly, because the spectrum
of an accretion disk -- being a superposition of black body spectra
-- is usually flatter than $S_{\nu}\propto \nu^2$. Only for high
inclination angles, nearly edge on, the disk spectrum comes close to a
black body type.  The fit of a one-component Planck function to a
flatter spectrum thus can lead to an underestimate of the actual
effective temperature of the inner edge of the disk. The uncertainties
in the observations still allow a broader range of spectral indices.

\section{The signature of rotating black~holes}
Fortunately the main results for standard accretion disks theory (von
Weizs\"acker \cite{L58}, Shakura \& Sunyaev
\cite{L5}) and its general relativistic extension (Novikov \& Thorne
\cite{L65}) follow from fundamental physics with only a few assumptions:

\begin{itemize}
\item The gravitational field is dominated by the
central black~hole and the metric is described by a Kerr metric (Kerr
\cite{L275}) allowing for rotation of the black~hole. Parameters of
the metric are mass $M$ and Kerr parameter $0 \le a\le 1$ of the black
hole ($a \propto$ angular momentum of black~hole).
\item Matter flows
along direct, nearly circular, geodesic orbits in the equatorial plane
of the black~hole.
\item Angular momentum is transported outwards due
to a turbulent viscosity mechanism resulting in a radial velocity of
the matter towards the center, which is much smaller than the circular
velocity of the matter.
\item The disk is thin $({\rm
height} < {\rm radius}/3)$ and steady.
\item At the inner edge of the
disk, angular momentum, mass and energy of the inflowing matter are
swallowed completely by the black~hole.
\end{itemize}

Having such an accretion disk the amount of gravitational energy which has
to be
dissipated per surface area and time (dissipation rate $D_0$) at each
 radius can
be calculated simply from conservation of energy and angular momentum,
yielding

\begin{equation} \relax\label{D0} D_{0}={{3GM \dot {M}} \over {8
\pi R^{3}}} {\cal {{Q} \over {B\sqrt{C}}}}=6.8 \cdot 10^{37}
{{ \dot {m}_{-8}} \over
{r^{3}}} \,  {\cal {{Q} \over {B\sqrt{C}}}}
\>
{\rm {{erg} \over {s
\,
{\mit R_{\rm g}^{2}}}}}
\end{equation}
which is {\em independent of the viscosity parameter $\alpha$} defined
by Shakura \& Sunyaev (1973). Assuming black-body radiation this can be
translated to an effective temperature via

\begin{equation}
\relax\label{Teff} T_{{\rm eff}}=\left( {D_{0}/\sigma }\right) ^{1/4}
\end{equation}
which in turn defines a characteristic frequency

\begin{equation}
\label{nuTeff}\nu _{{\rm max}}=8.2\cdot 10^{10}\> \left( {T_{{\rm eff}}/{\rm
K}}\right) \; {\rm Hz}
\end{equation}
where the black-body radiation of this temperature has a maximum
in $\nu F_\nu$:

\begin{equation}
\relax\label{numax} \nu _{{\rm max}}=7\cdot 10^{15}\> {{\dot {m}_{-8
}^{1/4}} \over {m_{6}^{1/2}}}\left( {{{1} \over {r^{3}}}{\cal
{{Q} \over {B\sqrt{C}}}}}\right) ^{1/4}{\rm Hz}\end{equation}

Here $m_6=M/10^6 M_{\sun}$ and $\dot{m}_{-8}=
\dot{M}/10^{-8}(M_{\sun}/{\rm yr})$ are the dimensionless mass and
accretion rate of the black~hole. \mbox{$r=R/R_{\rm g}$} is the dimensionless
radius in units of the gravitational radius $R_g=G M/c^2=1.48\cdot
10^5 M/M_{\sun} {\rm cm}$ which is half the Schwarzschild radius. The
inner edge of an accretion disk lies at the radius $r_{\rm ms}(a
\cdot R_{\rm g}$ of
the last, marginally stable orbit, which is $6\,R_{\rm g}$ for a
onrotating ($a=0$)
and $1.232\,R_{\rm g}$ for a maximal rotating black~hole ($a=0.9981$).
 The
relativistic correction factors ${\cal B,C}$ and ${\cal Q}$ are
functions of $r$ and $a$ with the limits such that $D_0 \propto
r^{-3}$ for $r\to\infty$ and $D_0\to 0$ for $r \to r_{\rm ms}$.  They
are given explicitly in Page \& Thorne (\cite{L63}). Note that the
dependence of the disk properties on the Kerr parameter $a$ is
contained completely in these correction factors and in the marginally
stable radius $r_{\rm ms}$. For fast rotating black~holes luminosity
and effective temperature of the disk may be up to $5$ times higher
than in the non-rotating case mainly because the disk extends to lower
radii. The radiation field of the disk is most sensitive to changes of
the holes angular momentum in the regime $a>0.9$.  Because of this
sensitivity one has for given black~hole mass and inclination angle  an
unequivocal translation from the two observables effective temperature
and luminosity $(T_{\rm eff},L)$ to a new pair of intrinsic
properties, namely angular momentum and accretion rate $(a,\dot{M})$
of the hole.

\section{Inner boundary conditions for a Kerr accretion disk}
Standard accretion disk theory assumes that at the inner boundary of
the accretion disk the effective temperature goes to zero. In the
stellar case a finite boundary layer has to exist which can produce an
appreciable part of the total luminosity, and which may also encompass
a non-negligible part of the accretion disk near the inner boundary
(Duschl \& Tscharnuter 1991).  In the case of black~hole accretion,
one expects a similar phenomenon (Kato 1982). But as
it is not necessary for the matter to slow down before being swallowed
by the hole as in the stellar case we neglect all such effects and use
the standard Novikov \& Thorne (1973) boundary condition described
at the beginning of Sect.~3. And even if there is a luminous
boundary layer most of the emitted photons will not be able to escape
from the very edge of the black~hole -- the relativistic
transferfunction tends to zero (Cunningham 1975), however the radial
structure of the disk may change in the inner part.

\section{Radiation field in the Kerr metric}
To build up a translation table from $(T_{\rm eff},L)$ to
$(a,\dot{M})$ we used the method from Cunningham (\cite{L207}) to
calculate the full general {rela\-ti\-vistic} geodesic transformation of
radiation in the Kerr-metric (i.e. gravitational redshift, boosting,
light bending). We assumed that the disk radiates locally as a
black-body with an additional limb-darkening law
$I_\nu(\theta)=(1+1.5\cos \theta)/1.75\cdot B_\nu $.  $B_\nu$~is the
Planck function with $T_{\rm eff}$ taken from Eqs.
(\ref{D0},\ref{Teff}). In our {sce\-nario}, where $M\sim 2\cdot 10^6
M_{\sun}$ and $\dot{M} \ll 10^{-5} M_{\sun}/{\rm yr}$, the standard
theory predicts an accretion disk which is dominated entirely by gas
pressure (Novikov \& Thorne \cite{L65}, Eqs. (5.9.6+8)) so that these
assumptions are justified with the exception of a possible weak
modification of the spectrum in the inner region of the disk due to
electron scattering, which was neglected. But since we are mainly
interested in the energetics of the disk radiation, we do not require
elaborate disk spectra at this stage.

Using the more detailed accretion disk theory (Novikov \& Thorne 1973)
-- including the visocsity paramter $\alpha$ -- one finds that the
disk is optically thick everywhere, for a low viscosity parameter
$\alpha\la 10^{-4}$ as used for the modeling of comparable disks for
FU Orionis events (Hessman 1991, Clarke et al. 1990). Only at low
accretion rates $\dot{m}\la 10^{-8} M_{\sun}/{\rm yr}$ and high
$\alpha$, an intermediate part of the disk may become optically thin.
However, this does not affect the outcome of our of the order
estimates drastically, as the high and low temperature parts of the
disk still remain optically thick.  The results do also not depend
crucially on the assumed limb-darkening law.

\begin{figure} \picplace{11.7cm} \caption[]{
\label{aTvsL}``Hertzsprung-Russell diagram for a black~hole'' of
$2\cdot 10^6 M_{\sun}$. Plotted is the effective temperature of the
radiation flux (maximum in $\nu F_\nu$) for a disk seen almost face on
(cos i= 0.9) and a disk seen almost edge on ($\cos i=0.05$) versus the
{\em total} luminosity of the disk for a grid of angular momenta and
accretion rates. The rectangle represents the observational
constraints from Sgr~A*. The arrow indicates how the whole grid shifts
if one divides the black~hole mass by a factor of 2.}
\end{figure}

 \begin{figure} \picplace{11.7cm} \caption[]{\label{Spectra}Sample
spectra of an accretion disk around a Kerr hole (a=0.9981) together
with the $\lambda=$2.2, 1.6 and 1$\mu$m data from Eckart et al., Close
et al. (1992) and Rosa et al. (1991). The {la\-bels} refer to different
 values of
mass and accretion rate. Spectra a)-d): face on disk with $M= (2\cdot
10^6, 2\cdot 10^5, 2\cdot 10^4, 2\cdot 10^3) M_{\sun}$ and
$\dot{M}=(10^{-10.5},10^{-9.5},10^{-8.5},10^{-7.5}) M_{\sun}/{\rm
yr}$.  Spectra e)-h): edge on disk with $M= (2\cdot 10^6, 10^6,
.5\cdot 10^6, .25\cdot 10^6) M_{\sun}$ and
$\dot{M}=(10^{-8.5},10^{-8.},10^{-7.5},10^{-7.}) M_{\sun}/{\rm yr}$.
The total luminosity for a maximal Kerr hole is given by $L_{\rm tot}
=2\cdot 10^{38} \dot{m}_{-8}$erg/s.}
\end{figure}

A variety of spectra were calculated as viewed by an observer at
infinity from different inclination angles for a black~hole with
$M=2\cdot 10^6 M_{\sun}$, angular momentum ranging from $a=0.1$ to
$a=0.9981$.  Because of relativistic boosting the effective
temperature of these spectra is maximal at high (edge on) and minimal
at low (face on) inclination angles, whereas the aparent luminosity is
lower at large and higher at low inclination angles. The temperature
effect is much more pronounced in the case of rotating black~holes
while the luminosity effect is stronger for non-rotating holes.

{}From each spectrum we took the maximum in $\nu F_\nu$ and converted
the corresponding frequency via Eq. (\ref{Teff}) to an effective
temperature. This corresponds to the effective temperature of the
energetically dominant part of the disk -- usually within a few
gravitational radii from the black~hole. The results are shown in Fig.
\ref{aTvsL} where we plotted these effective temperatures versus the $4
\pi$ integrated flux (i.e.  total luminosity) of the disk for a grid
of angular momenta and accretion rates. Also shown are our
observational constraints for the Galactic Center. Some selected
spectra for different parameters are shown in Figure \ref{Spectra}.

The spectral form of the disk radiation (exponential cut-off) ensures
that -- unless further hot components are present -- the effective
temperature is also very close to the high-energy cut-off for the
overall photon spectrum. That is very important vor the impact of the
disk radiation on the ambient medium, i.e. ionization and heating.
Different black~hole/accretion disk models may lead to different
temperatures in the surrounding gas due to photoionization. This
diagnostic tool will be considered in more detail in a subsequent
paper (Falcke et al. 1992).

We took a canonical black~hole mass of $2\cdot 10^6 M_{\sun}$ as our reference
model but one can easily obtain a similar grid for different masses simply by
using the Relations (\ref{D0}) and (\ref{numax}). As one can see the luminosity
scales with $L=2\int_{r_{\rm ms}}\!\!\!\!\! D_0 2\pi dr \propto \dot{M}$ and is
independent of the black~hole mass $M$, while the effective temperature scales
with $T_{\rm eff} \propto \dot{M}^{1/4}/\sqrt{M}$. Thus one can obtain further
models for different parameters of $M$ and $\dot{M}$ by simple linear
extrapolation.

\section{Discussion} If the observational constraints could
be confirmed then Fig. \ref{aTvsL} requires a fast
rotating Kerr hole ($a>0.9$) with a fairly low accretion rate of
$\dot{M}< 10^{-7} M_{\sun}/{\rm yr}$ seen almost edge on.

In fact if we fit the disk spectra directly to the NIR spectrum than
an even lower accretion rate of $\dot{M}\sim 10^{-8.5} M_{\sun}/{\rm
yr}$ seems also possible (Fig.\ref{Spectra}e), giving also a lower
limit, because an even lower accreting disk would not be capable of
reproducing the NIR data. From this figure one can also see -- like in
the HR-diagram -- that a low mass black~hole with $M\ll 10^6$ (e.g.
$\sim 10 M_{\sun}$ as proposed by Sanders (1992)) and an edge on disk
fitted to this data would produce a luminosity, exceeding the limit
from the dust observations, whereas the face on disk models produce
spectra which are obviously too flat.

So if the central disk really is edge on, then it is suggestive to
assume that it is also roughly aligned with the galactic plane, with
the rotation axis perpendicular to it. This, however, can not be
confirmed by NIR observation, but has to await mm VLBI experiments to
prove non-spherical structure in Sgr~A* and thus to fix the
orientation of the disk in the observers plane.

If the constraint on the mass of the putative black~hole could be
relaxed then a lower black~hole mass could also fit into the
Hertzsprung-Russell diagram, with no particular constraint on the Kerr
parameter.  But during the evolution process of a black~hole it is
quite natural to expect an evolution towards strong rotation to a
maximum at $a\approx 0.9981$ (Bardeen \cite{L277}, Thorne
\cite{L206}). From Fig.\ref{Spectra} one can see that a maximal Kerr
Hole of $M=10^6$ would then require at least $10^{-8} M_{\sun}/{\rm
yr}$.

Therefore the \GC could become an important laboratory to study
general relativity and the Kerr metric.  This model gives some firm
predictions \begin{itemize} \item strong anisotropies of the effective
temperature of the radiation field for a strongly rotating black~hole
resulting in an anisotropic ionization of the ambient medium, and
\item strong anisotropies for the luminosity of the radiation field
for non-rotating black~holes.
\end{itemize}

The results are summarized in Fig. \ref{MMdot}, where we show accretion
 rate versus black~hole mass with the parameters Kerr {para\-me\-ter}
 and inclination angle. We obtained this grid by applying the scaling
 laws $\dot{M}\propto \tilde{L}/L \cdot \dot{M}_0$ and
 $M=(\tilde{T}/T_{\rm eff})^2 \cdot
  \sqrt{L/\tilde{L}}\cdot M_0$ where $\tilde{L}=\tilde{L}(a,i)$ and
$\tilde{T}=\tilde{T}(a,i)$
 are the calculated functions for luminosity and effective temperature
of the disk at arbitrary but fixed values of $M=M_0$
 and $\dot{M}=\dot{M}_0$. $T_{\rm eff}$ and $L$ are the observed
 values of the radiation field  for which we took $T_{\rm eff}=4\cdot
 10^4 {\rm
  K}$ and $L=3\cdot 10^5 L_{\sun}$ as our reference frame for the
Galactic Center. Again, with
  better observational data at hand, one can easily adapt this grid by
 simple interpolation to test the black~hole scenario for Sgr~A*.

\begin{figure} \picplace{6.25cm} \caption[]{ \label{MMdot}
The shaded area limits the possible parameters $M$ and $\dot{M}$ for a
black~hole that matches a luminosity of $L=3\cdot 10^5 L_{\sun}$ and
an effective temperature $T_{\rm eff}=4\cdot 10^4 {\rm K}$ for a grid
of angular momenta $a$ and inclination angles $i$. The thick arrows
indicate how the whole grid shifts if one increases the luminosity by
a factor 2 or reduces the effective temperature by one tenth
respectively.}
\end{figure}

\acknowledgements{
HF is supported by the DFG (Bi 191/9). We thank Reinhold Schaaf and an
anonymous referee for helpful comments during preparation and revision
of this manuscript.}

\end{document}